\documentclass[aps,pre,twocolumn,showpacs,notitlepage,floatfix,superscriptaddress]{revtex4-1}
\usepackage{amssymb}
\usepackage{amsmath}
\usepackage{graphicx}
\usepackage{color}

\usepackage[all]{xy}
\input xy
\xyoption{matrix}
\xyoption{2cell}
\xyoption{arrow}
\xyoption{curve}
\xyoption{all}
\usepackage{graphicx}
\usepackage{subfigure}

\begin{document}
\title{
Compact Discrete Breathers on Flat Band Networks \\
}

\author{C. Danieli}
\affiliation{Center for Theoretical Physics of Complex Systems, Institute for Basic Science, Daejeon, Korea}
\author{A. Maluckov}
\affiliation{Vinca Institute for Nuclear Sciences, University of Belgrade, Serbia}
\affiliation{Center for Theoretical Physics of Complex Systems, Institute for Basic Science, Daejeon, Korea}
\author{S. Flach}
\affiliation{Center for Theoretical Physics of Complex Systems, Institute for Basic Science, Daejeon, Korea}
\affiliation{New Zealand Institute for Advanced Study, Massey University, Auckland, New Zealand}


\date{\today}

\begin{abstract}

Linear wave equations on flat band networks host compact localized eigenstates (CLS).
Nonlinear wave equations on translationally invariant flat band
networks can host compact discrete breathers - time-periodic and spatially compact localized
solutions. Such solutions can appear as one-parameter families of continued linear compact eigenstates,
or as discrete sets on families of non-compact discrete breathers, or even on purely dispersive networks with
fine-tuned nonlinear dispersion. In all cases, their existence relies on destructive interference. 
 We use CLS amplitude distribution properties and orthogonality conditions to derive
existence criteria and stability properties for compact discrete breathers as continued CLS.

\end{abstract}

\maketitle

\section*{Introduction}

In recent years, flat band tight binding networks gained interest in the fields of 
ultra cold atomic gases, condensed matter and photonics, among others \cite{leykam2018artificial}.
One of the essential features of the corresponding eigenvalue problem of these linear wave 
equations is the presence 
of eigenstates which are strictly compact in space.
These modes are coined {\it compact localized states}
(CLS), and their existence is due to destructive interference which
suppresses the dispersion along the network. The CLS
introduce macroscopic degeneracy in the energy spectrum of the
network, which results in one (or more) momentum independent (or
{\it dispersionless}) bands in the spectrum, hence called {\it flat}
bands. The CLS can be found irrespective to the dimensionality of
the network. 
CLSs can be classified according to the number $U$ of unit cells they occupy. 
Class $U = 1$ CLSs form an orthogonal basis of the flat band Hilbert space, since the compact states do not overlap. Moreover, the flat band can be freely tuned to be gapped away from dispersive bands, or to resonate with them.
Class $U \geq 2$ CLSs instead typically form a non-orthogonal basis, and the flat band is gapped away (or at most touching) from dispersive bands.

Introduced by Sutherland \cite{sutherland1986localization} and Lieb \cite{lieb1989two} in the 1980's,
and then generalized by Mielke and Tasaki in the
1990's \cite{mielke1991ferromagnetism,tasaki1992ferromagnetism},
flat band lattices and their perturbations provide an ideal
test-bed to explore and study unconventional localization and
innovative states of matter
\cite{bergholtz2013topological,derzhko2015strongly,moessner2006geometrical}.
The effects of different types of perturbations have been
studied in several examples of flat band networks
\cite{bodyfelt2014flatbands,danieli2015flatband},
as well as the effects of disorder and nonlinearity and
interaction between them \cite{Leykam2013Flat}. Further studies
focused on non-Hermitian flat band networks
\cite{Leykam2017Flat}, topological flat Wannier-Stark bands
\cite{kolovsky2018topological}, Bloch oscillations
\cite{khomeriki2016landau}, Fano resonances \cite{ramachandran2018fano},
fractional charge transport \cite{flach2017fractional} and the existence of nontrivial
superfluid weights \cite{peotta2015superfluidity}. Chiral flat
band networks revealed that CLS and their macroscopic degeneracy
can be protected under any perturbation which does not lift the
bi-partiteness of the network \cite{Ramachandran2017chiral}. The
engineering of CLS has been longly attempted
\cite{dias2015origami,morales2016simple}, and it has been recently
solved for $U=1$ lattices \cite{flach2014detangling} and for
the $U=2$ CLS in a two-band problem \cite{maimaiti2017compact}.
Experimentally, compact localized states have been realized using
ultra cold atoms \cite{taie2015coherent}, photonic waveguides
networks
\cite{vicencio2015observation,mukherjee2015observation,weimann2016transport}
exciton-polariton condensates
\cite{masumoto2012exciton,baboux2016bosonic} and superconducting
wires \cite{vidal1998aharonov,abilio1999magnetic} (for a recent
survey on the state of the art, see Ref.\cite{leykam2018artificial}).

Nonlinear translationally invariant lattices admit a class of time-periodic solutions localized in real space
(typically exponentially), called {\it discrete breathers} \cite{flach1998discrete,flach2008discrete}. 
The precise decay in the tails depends on the band structure of small amplitude linearized wave equations. For analytic band structures (usually due to short range - 
e.g. exponentially or faster decaying -
connectivities on the lattice), 
the discrete breather tails decay exponentially. For non-analytic band structures (usually due to long range - e.g. algebraically decaying - connectivities on the lattice),
the tails decay algebraically as well. In the absence of linear dispersion, but presence of nonlinear dispersion, tails decay superexponentially. For short range connectivities, but with acoustic parts in the
band structure, and with broken space parity, the ac parts of the discrete breather tails decay exponentially, while the dc part (static lattice deformation) will
decay algebraically \cite{flach1998discrete,flach2008discrete}.

A natural question then arises whether discrete breathers can have strictly zero tails, and turn into compact excitations. 
For instance, traveling solitary waves with compact support have been found in the frame of spatially continuous partial differential equations by Rosenau and Hyman in the Korteweg-deVries model \cite{rosenau1993compactons}. 
In discrete systems, 
spatially compact time-periodic solutions have been found by Page in a purely anharmonic one-dimensional {\it Fermi-Pasta-Ulam}-like chain in the limit of non-analytic compact (box) interaction potential \cite{page1990asymptotic}.
Moreover, Kevrekidis and Konotop reported on compact solutions in translationally invariant one-dimensional lattices in the presence of non-local nonlinear terms \cite{kevrekidis2002bright}.
In this work, we consider flat band networks as the underlying support for compact time-periodic excitations.

The existence of compact discrete breathers in nonlinear flat band networks  
was observed in 
Ref.
\cite{johansson2015compactification,belivcev2017localized}. 
Furthermore, the coexistence between nonlinear terms and
spin-orbit coupling has been discussed in the framework of
ultra-cold atoms in a diamond chain
\cite{gligoric2016nonlinear}. 
Perchikov and Gendelman studied compact time-periodic solutions in a one-dimensional nonlinear mechanical cross-stitch network \cite{perchikov2017flat}.
In this case the above mentioned destructive interference translates into several time-dependent forces
acting on masses in the mechanical network in such a way that the sum of all forces vanishes, leading
to a compactification of the vibrational excitation.

In this work, we present a necessary and sufficient condition for the existence and continuation of time-periodic and compact in
space solutions (herewith called {\it compact discrete breather}) 
on flat band networks with local nonlinearity. 
The existence and continuation condition applies irrespective of the
dimensionality of the lattice and the class $U$ of linear CLS.
Then, we discuss the linear stability of compact discrete breathers.
For orthogonal CLSs in $U=1$ networks, the only source of instability are resonances
with extended states.
For class $U\geq 2$ networks instead, the non-orthogonality between linear CLSs induces additional potential local instabilities due to CLS-CLS interaction. 
Resonances with dispersive states lead to radiation and potential complete annihilation.
Resonances with neighboring CLSs in general simply yield local instabilities which do not annihilate 
the excitation.
The study of the nonlinear stability has been performed numerically, and
standard techniques of perturbation theory have been applied to
substantiate the numerical findings.
The present work is structured as follows: in Sec.\ref{sec:FB}
we will present the flat band networks;  then in
Sec.\ref{sec:NLFB} we introduce the nonlinear terms in flat band model
equations, and discuss the continuation criteria of linear
CLS to compact discrete breathers. Next, in Sec.\ref{sec:LSA_a} we
present the linear stability analysis of the compact discrete
breathers, which will then be discussed numerically
in Sec.\ref{sec:LSA_num}.

\section{Flat Band Networks}\label{sec:FB}

For simplicity we will operate in one spatial dimension. Results in general take over to higher dimensions. We will comment on particular cases
where caution is to be executed. The linear time-dependent model equation of the flat band networks can be presented in a form
\begin{equation}
i \dot{\psi}_n = H_0\psi_n + H_1\psi_{n+1} + H_1^\dagger\psi_{n+1}
\label{eq:FB_ham1}
\end{equation}
For all $n\in\mathbb{Z}$, $\psi_n \in\mathbb{C}^\nu$ is a time-dependent complex vector of $\nu$ components, each one representing one site of the network. The set of $\nu$ sites is called the {\it unit-cell}. The matrix $H_0$ defines the geometry of the unit-cell, while the matrixes $H_1, H_1^\dagger$ define the hopping between nearest-neighboring ones. This model equation can be easily generalized to longer range hopping as well as higher dimensional networks.
The phase-amplitude ansatz $\psi_n(t) = A_n e^{-i E t}$ leads to the associated eigenvalue problem
\begin{equation}
E A_n = H_0 A_n + H_1 A_{n+1} + H_1^\dagger A_{n+1}
\label{eq:EV_FB_ham1}
\end{equation}
Then, the Bloch solution $A_n = \varphi_q e^{i q n }$ of Eq.(\ref{eq:EV_FB_ham1}) defined for the wave-vector $q$ gives rise to the Bloch hamiltonian of the lattices
\begin{equation}
E \varphi_q = H (q)  \varphi_{q} \equiv \big[ H_0 + e^{i{ q}}H_1 + e^{-i{ q}} H_1^\dagger \big] \varphi_{ q}
\label{eq:EV_FB_ham2}
\end{equation}
Eq.(\ref{eq:EV_FB_ham2}) yields the band structure $E = \cup_{i=1}^\nu E_i( {q})$ of the problem. We consider lattices which exhibit at least one band independent from the wave vector $q$, which we call disperionless (or {\it flat}) band $E_{FB}$. The eigenmodes associated to a flat band are typically compact localized states (CLS), and the number $U$ of unit-cells occupied by one CLS is the flat band {\it class}. These states can be written in the time-dependent form solutions of Eq.(\ref{eq:FB_ham1})
\begin{equation}
\begin{split}
 {\bf \psi}_{n,n_0} (t) &= \Bigg[ \sum_{l=0}^{U-1} v_l \delta_{n,n_0 + l}\Bigg] e^{-i E_{FB} t}
  \end{split}
\label{eq:FB_states1}
\end{equation}
where the sum indicates the spatial component of the CLSs. The real vectors $v_l$ are defined as the following
\begin{equation}
\begin{split}
 &  v_l=  \sum_{j=1}^{\nu} a_{l,j} A_{l,j} {\bf e}_j
 \end{split}
\label{eq:v_FB_states1}
\end{equation}
where $\{ {\bf e}_j \}_{j=1}^\nu$ are the canonical basis of $\mathbb{R}^\nu = \langle {\bf e}_1,\dots, {\bf e}_\nu \rangle$;
$a_{l,j} \in \{ 0,\pm 1 \} $ denotes the sites with non-zero amplitude, and the real numbers  $A_{l,j} \in\mathbb{R} $ defines the  amplitudes in the sites with non-zero $a_{i,j}$.
In the next section, we will introduce local nonlinear terms to Eq.(\ref{eq:FB_ham1}), and we will discuss continuation criteria for the CLS introduced in Eq.(\ref{eq:FB_states1}) as compact solution of the nonlinear regime.

\section{Nonlinear Flat Band Networks and Continuation of Compact Localized States}\label{sec:NLFB}

Let us consider the model equation of the flat band network Eq.(\ref{eq:FB_ham1}) in presence of local nonlinear terms
\begin{equation}
\begin{split}
i \dot{\psi}_n = H_0 \psi_n &+ H_1\psi_{n+1} + H_1^\dagger\psi_{n+1} + \gamma  \mathcal{F}( \psi_n) \psi_n
 \end{split}
\label{eq:FB_ham_NL1}
\end{equation}
where the matrix $\mathcal{F}( \psi_n)$
\begin{equation}
\begin{split}
&  \mathcal{F}( \psi_n) \equiv \sum_{j=1}^\nu |\psi_n^j|^{2} \ {\bf e}_j\otimes {\bf e}_j
 \end{split}
\label{eq:FB_ham_NL1_2}
\end{equation}
contains the terms $|\psi_n^i|^{2}$ along the diagonal. We seek for time-periodic solutions of the nonlinear system Eq.(\ref{eq:FB_ham_NL1})
\begin{equation}
\begin{split}
 C_{n,n_0} (t) &= \Bigg[ \sum_{l=0}^{U-1} v_l \delta_{n,n_0 + l}\Bigg] e^{-i \Omega t} 
 \end{split}
\label{eq:NL_FB_states1}
\end{equation}
with frequency $\Omega$, which are continuation of the CLSs Eq.(\ref{eq:FB_states1},\ref{eq:v_FB_states1}) that exist in the linear regime $\gamma = 0$.

We consider the compact solution Eq.(\ref{eq:NL_FB_states1}) defined with the profile in space of the linear CLS in Eq.(\ref{eq:FB_states1},\ref{eq:v_FB_states1}) and frequency $\Omega$, and check under which conditions these are solutions of the nonlinear equation Eq.(\ref{eq:FB_ham_NL1}). At first, let us observe that for all sites where a CLS is zero ($a_{l,j} = 0$ and outside the range of $U$ cells in Eq.(\ref{eq:FB_states1})), Eq.(\ref{eq:FB_ham_NL1}) is solved.
For $l = 1,\dots ,U$ and $j =1, \dots, \nu$ where a CLS has non-zero amplitude $a_{l,j} \neq  0$, Eq.(\ref{eq:FB_ham_NL1}) reduces to
\begin{equation}
\Omega A_{l,j} = E_{FB} A_{l,j} + \gamma A_{l,j}^{3} \ .
\label{eq:NL_FB4}
\end{equation}
If for all $l,j$ such that $a_{l,j} \neq  0$, $|A_{l,j}| \equiv A$ (all sites have same amplitude in absolute value), Eq.(\ref{eq:NL_FB4}) turns into
\begin{equation}
 A^{2} = \frac{\Omega - E_{FB}}{\gamma} \ ,
\label{eq:NL_FB4_b}
\end{equation}
If instead there exist non-zero $|A_{l,j} | \neq |A_{\hat{l}, \hat{j}}|$, Eq.(\ref{eq:NL_FB4}) yields different frequencies $\Omega$, which breaks the condition of continuation of CLS as a periodic orbit with compact support. Let us introduce the following definition:\\

{\bf Definition}: let  ${\bf \psi}_{n,n_0} (t)$ be a CLS of class $U$ of a flat band network with $\nu$ sites per unit-cell.
We call ${\bf \psi}_{n,n_0}(t)$ a {\it homogeneous} CLS if
\begin{equation}
\text{for all} \ \ a_{l,j} \neq 0 \quad\Rightarrow\quad  |A_{l,j}| \equiv A
\label{eq:CDB_def1}
\end{equation}
and we call ${\bf \psi}_{n,n_0}(t)$ a {\it heterogeneous} CLS otherwise.\\
\\
From the above consideration in Eq.(\ref{eq:NL_FB4},\ref{eq:NL_FB4_b}), we can obtain the following continuation criteria in the following lemma\\

{\bf Lemma}: in a nonlinear flat band network Eq.(\ref{eq:FB_ham_NL1}), a compact state ${\bf \psi}_{n,n_0} (t)$ of the linear lattice
$\gamma=0$ with energy $E_{FB}$ can be continued as a periodic
orbit with compact support $C_{n,n_0} (t)$ with frequency $\Omega = E_{FB} +
\gamma A^{2}$
{\it if and only if}
it is homogeneous.\\
%
%

This lemma states a necessary and sufficient condition for linear CLSs to be continued as time-periodic solutions of the nonlinear regime with compact support.  Indeed, {\it homogeneous} CLSs in presence of this local nonlinearity do not break the destructive interference, preserving therefore the compactness in space. {\it Heterogeneous} CLSs instead in presence of nonlinearity break the destructive interference, loosing therefore the compactness in space.
We call the continued homogeneous CLS solutions {\it compact discrete breathers}. 
Their spatial profile is identical to the CLS one, and their frequency is given by
\begin{equation}
\Omega = E_{FB} + g \;,\qquad g\equiv \gamma A^2\;.
\label{dbfrequency}
\end{equation}
In the next section, we will discuss the linear stability of compact discrete breathers.

\section{Linear Stability Analysis}\label{sec:LSA_a}

Herewith, we consider a perturbation $ \epsilon_n(t)$ of a compact discrete breather $C_{n,n_0} (t)$ Eq.(\ref{eq:NL_FB_states1}) solution of the nonlinear flat band model Eq.(\ref{eq:FB_ham_NL1})
\begin{equation}
  \begin{split}
  &\psi_n (t) =  C_{n,n_0} (t) + \epsilon_n(t)\ ,\\
  \end{split}
\label{eq:CLS_perturbed1}
\end{equation}
By linearizing Eq.(\ref{eq:FB_ham_NL1}) around one compact discrete breather $C_{n,n_0} (t)$,
and defining $g \equiv \gamma A^2 $, we obtain
\begin{equation}
\begin{split}
i \dot{\epsilon}_n &=
  H_0\epsilon_n + H_1\epsilon_{n+1} + H_1^\dagger\epsilon_{n+1} \\
  & + g  \sum_{l=0}^{U-1} \Gamma_l \big(2 \epsilon_{n} + e^{- i 2 \Omega t} \epsilon_{n} ^*   \big) \delta_{n,n_0 + l}
\end{split}
\label{eq:FB_ham_NL3}
\end{equation}
where $\{ \Gamma_l \}_{l=0}^{U-1}$ are the projector operators of $\psi_n$ over a compact discrete breather $C_{n,n_0} (t)$
\begin{equation}
 \Gamma_l  = \sum_{j=1}^\nu a_{l,j} {\bf e}_j \otimes {\bf e}_j
\label{eq:projector}
\end{equation}
The resulting dynamical model Eq.(\ref{eq:FB_ham_NL3}) for the perturbation term $\epsilon_n$ consists of equations with time-dependent coefficients that occur 
at sites where the compact discrete breather  $C_{n,n_0} (t)$ has non-zero amplitudes. The aim of this section is to analytically prove the existence of regions of instability 
in the parameter space $(\Omega, g) \in\mathbb{R}\times\mathbb{R}$ for the compact discrete breather. In order to achieve this, we first express Eq.(\ref{eq:FB_ham_NL3}) in the Bloch representation. Than, we compute the condition for resonance determining the Floquet matrix at $g=0$. At last, we obtain the regions of instability around the resonances via the strained coefficient method, focusing on the $U=1$ and $U=2$  cases.

\subsection{Bloch States Representation}\label{sec:Bloch}

Let us consider the Bloch representation of Eq.(\ref{eq:FB_ham_NL3}) using the following transformation
\begin{equation}
\epsilon_n = \frac{1}{\sqrt{N}} \sum_{q} \phi_q e^{iqn}
\label{eq:bloch_expansion}
\end{equation}
This leads to the Bloch equation,
\begin{equation}
\begin{split}
&\qquad \qquad \qquad  i \dot{\phi}_{\hat{q}} =  H(\hat{q})\phi_{\hat{q}} \\
&+ \frac{ g }{N}  \sum_{q}  \Bigg[ \sum_{l=0}^{U-1} e^{-i\hat{q}l} \Gamma_l
\bigg(2 e^{iq l} \phi_q + e^{- i 2 \Omega t}  e^{-i q l } \phi_q^*  \bigg)  \Bigg]
\end{split}
\label{eq:FB_ham_bloch2}
\end{equation}
where $H(q) \equiv  H_0 + e^{iq}H_1+ e^{-iq}H_1^\dagger$ is the Bloch matrix.
The $H(q)$ matrix admits $\nu$ eigenvectors $v_q^i$ and $\nu $ eigenvalues $\lambda_q^i$.
We assume that one flat band $\lambda_q^1 = E_{FB}$ exists with
corresponding eigenvector $w_q$ of the Bloch matrix.
Then we define the expansion of $\phi_{\hat{q}} $ in the Bloch eigenbasis
\begin{equation}
\begin{split}
\phi_{\hat{q}} = f_{\hat{q}} w_{\hat{q}}  +  \sum_{i=2}^\nu d_{\hat{q}}^i v_{\hat{q}}^i
\end{split}
\label{eq:bloch_exp1}
\end{equation}
The resulting equations on the expansion coefficients $ f_{\hat{q}}$ of the flat band reads (see Appendix \ref{sec:app_1_bloch}) 
\begin{equation}
\begin{split}
&\qquad\qquad\quad  i \dot{f}_{\hat{q}}   =
  E_{FB}  f_{\hat{q}}  \\
  &+ \frac{ g }{N}  \sum_{q}  \Bigg\{ \sum_{l=0}^{U-1} e^{-i\hat{q}l}
  \Big( 2 e^{iq l} f_q + e^{- i 2 \Omega t}  e^{-i q l } f_q^* \Big) \Gamma_l w_q \cdot w_q^* \\
& + \sum_{l=0}^{U-1}  \Bigg[ \sum_{i=2}^\nu  e^{-i\hat{q}l} \Big( 2 e^{iq l} d_q^i + e^{- i 2 \Omega t}
  e^{-i q l } d_q^{i *} \Big) \Gamma_l v_q^i \cdot w_q^* \Bigg] \Bigg\}
\end{split}
\label{eq:FB_ham_bloch_DB}
\end{equation}
while the equation of the coefficients $d_{\hat{q}}$ of the $j$-th dispersive band reads
\begin{equation}
\begin{split}
&\qquad\qquad\quad  \dot{d}_{\hat{q}}^j =
 \lambda_q^j d_{\hat{q}}^j  \\
 &+ \frac{ g }{N}  \sum_{q}  \Bigg\{ \sum_{l=0}^{U-1} e^{-i\hat{q}l} \Big( 2 e^{iq l} f_q + e^{- i 2 \Omega t}  e^{-i q l } f_q^* \Big) \Gamma_l w_q \cdot v_q^{j *} \\
& + \sum_{l=0}^{U-1}  \Bigg[ \sum_{i=2}^\nu  e^{-i\hat{q}l} \Big( 2 e^{iq l} d_q^i + e^{- i 2 \Omega t}  e^{-i q l } d_q^{i *} \Big) \Gamma_l v_q^i \cdot v_q^{j *} \Bigg] \Bigg\}
\end{split}
\label{eq:FB_ham_bloch_DB2}
\end{equation}

Eq.(\ref{eq:FB_ham_bloch_DB}) and Eq.(\ref{eq:FB_ham_bloch_DB2}) describe the time-dynamics of the flat band
states $f_q$ with  dispersive states $d_q^i$ due to the linearized term of Eq.(\ref{eq:FB_ham_NL3}). For class $U=1$, these equations are
decoupled, while for class $U>1$ they are coupled. In the next subsection, we neglect the terms following from the nonlinearity (set $g=0$), and obtain the resonance condition by computing the Floquet matrix of the system Eq.(\ref{eq:FB_ham_bloch_DB},\ref{eq:FB_ham_bloch_DB2})

\subsection{Floquet Matrix}\label{sec:floquet}

For $g=0$, we calculate the {\it Floquet matrix} $A$ for Eq.(\ref{eq:FB_ham_bloch_DB},\ref{eq:FB_ham_bloch_DB2}) (also called {\it period advancing matrix}).
For $\varphi = (f_q , d_q)$ and $T = \pi / \Omega$, it follows that
\begin{equation}
      \begin{bmatrix}
         \varphi(t + T)   \\
         \varphi^*(t + T)
      \end{bmatrix} 
      =A
     \begin{bmatrix}
         \varphi(t )   \\
         \varphi^*(t )
          \end{bmatrix}
          \qquad \text{with}\qquad
          A \equiv
             \begin{bmatrix}
        e^{-i\lambda T} & 0  \\
        0 & e^{i\lambda T}
              \end{bmatrix}
\label{eq:FM1}
\end{equation}
The eigenvalues of the Floquet matrix $A$ will be degenerate on the unit circle {\it if and only if} $\cos^2(\lambda T) = 1$, which means
\begin{equation}
\begin{split}
\cos^2(\lambda T) = 1 &\quad\Leftrightarrow\quad  \lambda T = m\pi \ ,\quad m\in\mathbb{Z} \\
&\quad\Leftrightarrow\quad  \lambda = m\Omega  \ ,\quad m\in\mathbb{Z}
\end{split}
\label{eq:FM_res}
\end{equation}
Concerning Eq.(\ref{eq:FB_ham_bloch_DB},\ref{eq:FB_ham_bloch_DB2}), Eq.(\ref{eq:FM_res}) implies that for $m\in\mathbb{Z} $
\begin{equation}
\begin{split}
E_{FB} &= m\Omega \ ,\quad
 \lambda_q^j = m\Omega  \ ,\quad  j=2,\dots,\nu
\end{split}
\label{eq:FM_res_2}
\end{equation}
It follows that from the values  of the frequency $\Omega$ of a compact discrete breather $C_{n,n_0} (t)$ contained in Eq.(\ref{eq:FM_res_2}),
regions of instability (Arnol'd tongues) in the parameter space $(\Omega,g)$ are expected. In order to obtain an approximation of these regions, we apply a standard technique of perturbation theory called {strained method coefficient}.

\subsection{Arnol'd tongues}\label{arnold}\label{sec:arnold_tongue}

In the following, we estimate the regions of instability in the parameter space $(\Omega,g)$ of
Eq.(\ref{eq:FB_ham_bloch_DB}) and Eq.(\ref{eq:FB_ham_bloch_DB2}), separating between the class $U=1$ case (where the dispersive states are decoupled from the flat band ones) and the class $U=2$ case (where dispersive and flat band states are coupled).

\subsubsection{Class $U=1$} \label{sec:arnold_tongue_U=1}

\begin{figure}[h]
\includegraphics [width=0.8\columnwidth]{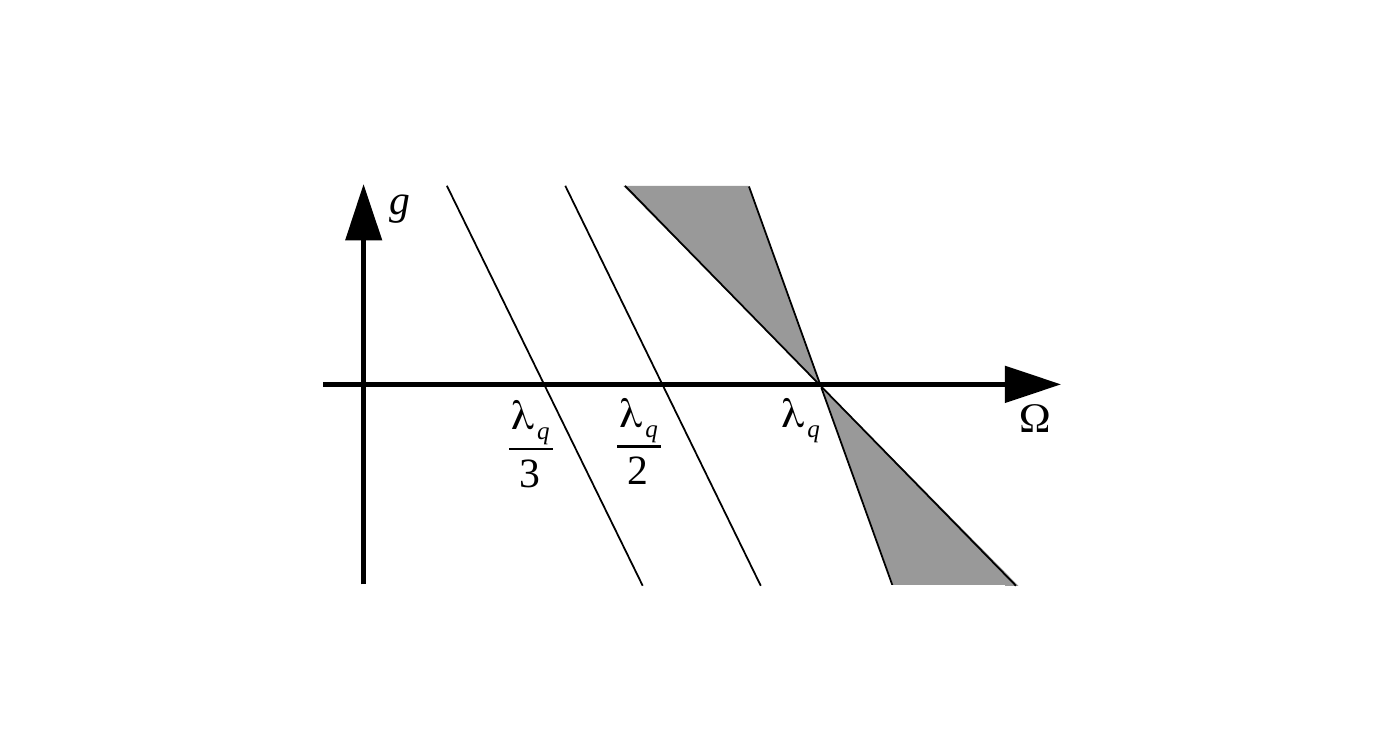}
\caption{First order approximation of the Arnold's tongues (grey shaded area) at a dispersive energy $\lambda_q$.} \label{fig:arnoldU1}
\end{figure}

In the case of class $U=1$ flat band network, it holds that $\Gamma_0 w_q = w_q$ and Eq.(\ref{eq:FB_ham_bloch_DB},\ref{eq:FB_ham_bloch_DB2}) reduce to
\begin{equation}
\begin{split}
 i  \dot{f}_{\hat{q}}  & =
  E_{FB}  f_{\hat{q}}  + \frac{ g }{N}  \sum_{q}  \Big( 2 e^{iq l} f_q + e^{- i 2 \Omega t}  e^{-i q l } f_q^* \Big)  \\
i \dot{d}_{\hat{q}}^j& =
 \lambda_q^j d_{\hat{q}}^j  + \frac{ g }{N} \sum_{i=2}^\nu   \Bigg\{  \sum_{q}
  2 d_q^i + e^{- i 2 \Omega t}  d_q^{i *}  \Bigg\}  \Gamma_0 v_q^i \cdot v_q^{j *}
\end{split}
\label{eq:FB_ham_bloch_DB_U=1}
\end{equation}
Without loss of generality, we refer to a two bands problem $\nu = 2$. The equations of the dispersive band component $d_q$ of Eq.(\ref{eq:FB_ham_bloch_DB_U=1}) read
\begin{equation}
\begin{split}
i \dot{d}_{\hat{q}}& =
 \lambda_q d_{\hat{q}} + \frac{ g }{N}  \sum_{q}
 \Big( 2 d_q + e^{- i 2 \Omega t}  d_q^{ *} \Big)
\end{split}
\label{eq:FB_ham_bloch_DB_U=1_CS}
\end{equation}
The strained coefficient method consists in expanding in powers of $g$ both the time-dependent component $d_q$ as well as the frequency term $\lambda_{\hat{q}}$ around one of the resonant frequencies in Eq.(\ref{eq:FM_res_2}). Then, we determine the expansion coefficients so that the resulting expansion is periodic. This will define transition curves between stability and instability regions in the parameter space $(\Omega,g)$ (for further details, see \cite{nayfeh1993introduction}).
The expansion of $d_q$ and $\lambda_{\hat{q}}$ reads
\begin{equation}
\begin{split}
d_{\hat{q}} &=\sum_{k=0}^{+\infty} g^k u_k^{(\hat{q})} \ , \quad
\lambda_{\hat{q}} = m\Omega + \sum_{l=1}^{+\infty} g^l \delta_l  
\end{split}
\label{eq:nayfeh_exp1}
\end{equation}
for $\lambda_{\hat{q}} \neq \lambda_q$ and for all $q\neq \hat{q}$. Vanishing the {\it secular terms} (terms which give rise to non-periodicities in the expansion) demands the following conditions in the expansion coefficient $\delta_1$ (see Appendix \ref{sec:app_2_strained} for details)
\begin{equation}
\begin{split}
&\delta_1 = -\frac{3}{N},-\frac{1}{N}  \ ,\quad m=1\\
& \delta_1 = -\frac{2}{N} \ , \qquad \quad\  m\neq 1
\end{split}
\label{eq:CS_secular1_DB}
\end{equation}
This implies that that a region of instability appears from each dispersive frequency $\lambda_{q}$ (obtained for $m=1$ in Eq.(\ref{eq:FM_res_2})), while for $\lambda_{q} / m$ for $m\geq 2$ regions of instability are absent. It is important to notice that for $N\longmapsto \infty$ the coefficients in Eq.(\ref{eq:CS_secular1_DB}) 
converge to zero, implying that in the limit of infinite chain, the instability regions disappear, in analogy with \cite{marin1998finite}.
In Fig.\ref{fig:arnoldU1} we can see a representation of the Arnol'd tongue around one frequency $\lambda_{q}$.

Analogous conclusions follow from the strained coefficient method applied to the flat band states $f_q$. Here the expansions reads
\begin{equation}
\begin{split}
f_{\hat{q}} =\sum_{k=0}^{+\infty} g^k v_k^{(\hat{q})}\ ,\qquad E_{FB} = m\Omega + \sum_{l=1}^{+\infty} g^l \sigma_l 
\end{split}
\label{eq:nayfeh_exp_fb1}
\end{equation}
The zeroing of the secular terms yields to the following coefficients
\begin{equation}
\begin{split}
&\sigma_1 = -3,-1  \ ,\quad m=1\\
& \sigma_1 = -2 \ , \qquad \quad m\neq 1
\end{split}
\label{eq:CS_secular1_FB}
\end{equation}
Eq.(\ref{eq:CS_secular1_FB}) is independent on $N$ due to macroscopic degeneracy of the flat band states (see Appendix \ref{sec:app_2_strained} for details).

The strained coefficient method showed the appearance of regions of instability in correspondence of each dispersive energy $\lambda_{q}$ of the dispersive band. However, instability regions do not appear for higher order resonances ($\lambda_{q}/m$ for $m \geq 2$) (see Fig.\ref{fig:arnoldU1}).
Furthermore, we can notice that this  region of instability also follows from the Bogoliubov expansion of Eq.(\ref{eq:FB_ham_bloch_DB_U=1}) (see Appendix \ref{sec:app_3_Bogo} for details). Before to go ahead to numerical studies, we briefly check the previous approach
in the case with $U=2$.

\subsubsection{Class $U \geq 2$} \label{sec:arnold_tongue_U>1}

In the case $U=2$, without loss of generality, we refer to a two band problem $\nu=2$, using the saw-tooth network as a test-bed. Eq.(\ref{eq:FB_ham_bloch_DB},\ref{eq:FB_ham_bloch_DB2}) read (see Appendix \ref{sec:app_2_strained} for details)
%
 \begin{equation}
\begin{split}
 i \dot{f}_{\hat{q}}  =
  E_{FB}  f_{\hat{q}}  + \frac{ g }{ \alpha^2 N }  &\sum_{q}  \Bigg\{ (\alpha^2 - 1) \Big(  f_q + e^{- i 2 \Omega t}  f_q^* \Big)   \\
  &+ e^{-i\hat{q}} \Big( 2 e^{iq } f_q + e^{- i 2 \Omega t}  e^{-i q  } f_q^* \Big) \\
  &+    (1 + e^{-iq})  \Big( 2 d_q^i + e^{- i 2 \Omega t}   d_q^{i *} \Big)  \Bigg\}\\
  \end{split}
\label{eq:bloch_U=2_ST_fb}
\end{equation}
    \begin{equation}
\begin{split}
i \dot{d}_{\hat{q}} =
 \lambda_q d_{\hat{q}}
 + \frac{ g }{\alpha^2 N} & \sum_{q}  \Bigg\{
    \Big( 2 d_q + e^{- i 2 \Omega t}  d_q^{*} \Big)   \\
&+  e^{-i\hat{q}} \Big( 2 e^{iq } d_q+ e^{- i 2 \Omega t}  e^{-i q  } d_q^{*} \Big)  \\
&+  (1 + e^{iq})  \Big( 2 f_q + e^{- i 2 \Omega t}  f_q^* \Big)
\Bigg\}
\end{split}
\label{eq:bloch_U=2_ST_db}
\end{equation}
for $\alpha = \sqrt{3 + 2\cos q}$. Both expansions Eq.(\ref{eq:nayfeh_exp1},\ref{eq:nayfeh_exp_fb1}) have to be applied to Eq.(\ref{eq:bloch_U=2_ST_fb},\ref{eq:bloch_U=2_ST_db}). 
However, in the first order, the additional terms (the second and the third lines of both equations) do not provide the appearance of further regions of instability (see Appendix \ref{sec:app_2_strained} for details). These additional {\it polarized} terms (terms dependent on the wave number $q$) indeed provide interactions between dispersive and flat band states. However, the strained coefficient method does not report additional instability regions in the parameter space $(\Omega,g)$ due to these terms.
In the following, we will discuss numerically the linear stability of the compact discrete breather solutions of certain examples of class $U=1$ and class $U=2$ one-dimensional nonlinear flat band networks.

\section{Numerical Results}\label{sec:LSA_num}

In this section we numerically study the linear stability
properties of the compact discrete breather solutions of certain
flat band topologies. We then relate the numerical observations
with the analytical results discussed above. Herewith we
numerically solve the eigenvalue problem Eq.(\ref{eq:FB_ham_NL3})
obtained from the time-evolution equations
Eq.(\ref{eq:FB_ham_NL1}) linearized around a compact discrete
breather Eq.(\ref{eq:CLS_perturbed1}). Generally, we will obtain
complex eigenvalues, and the presence of non-zero real part will
highlight instability \cite{lichtenberg1992regular}. 
We will also discuss the
nature of the eigenvector associated to unstable eigenvalues
(eigenvalues with non-zero real part). Furthermore, we will
substantiate the findings by showing simulations of the time
evolution of initially perturbed Compact Discrete Breathers. In
the following, we will focus on two models: the
{\it Cross-Stitch} lattice and the {\it Saw-Tooth} chain.
In Appendix \ref{sec:app_4_nummeth} we detail the numerical methods used along the work.

\subsection*{Cross-Stitch Lattice}

\begin{figure}[h]
\centering
\includegraphics [width=\columnwidth]{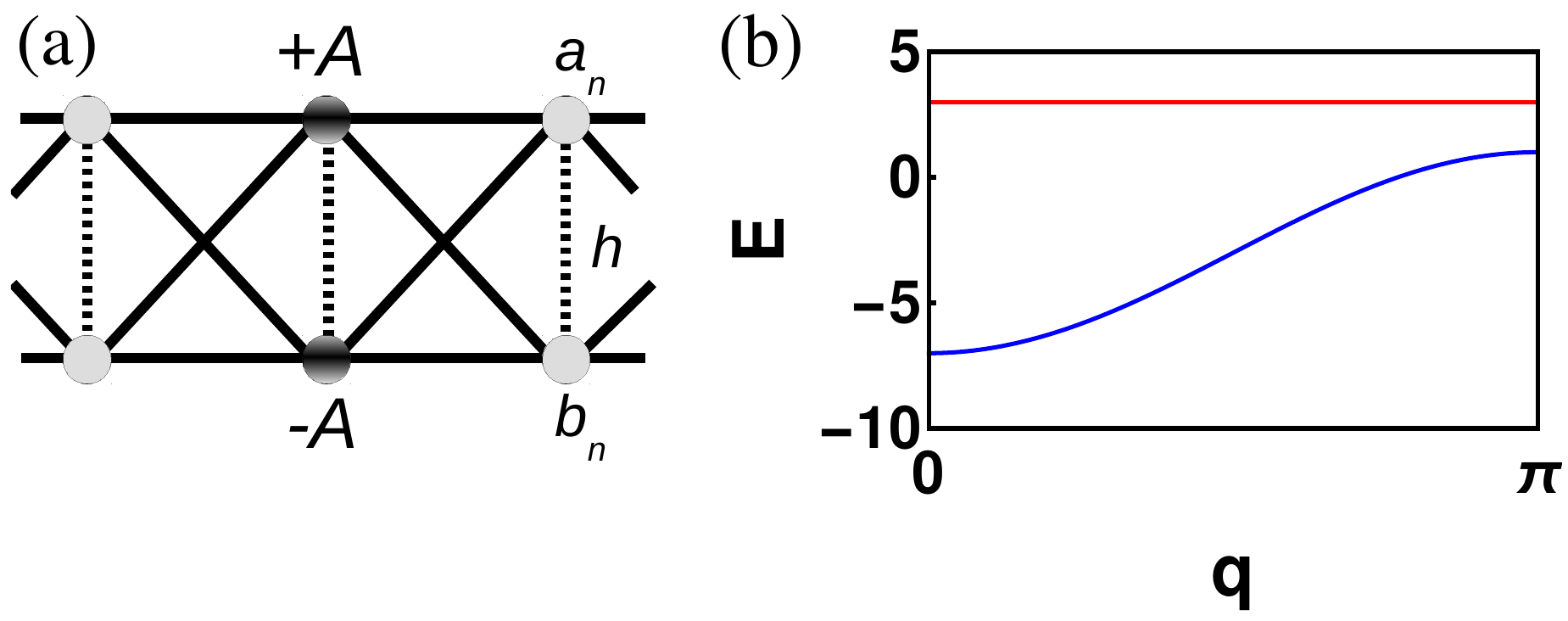}
\caption{(a): profile of the
Cross-Stitch lattice. (b): band structure for $h=3$.}
\label{figCS1}
\end{figure}

The cross-stitch lattice - Fig.\ref{figCS1}(a) - is a
one-dimensional two-band network, which posses one flat band.
Associated to the flat band, there exists a countable set of class
$U=1$ compact localized states, whose homogeneous profile in space
is shown by the black dots in Fig.\ref{figCS1}(a). The full band
structure of the model is
\begin{equation}
E_{FB} = h\ ,\qquad E(q) = -h + 4\cos(q)
\label{eq:CS_spectrum1}
\end{equation}
which can be visualized in Fig.\ref{figCS1}(b),
for $h=3$.
In this model, the relative position between dispersive and flat bands can be tuned using the free parameter $h\in\mathbb{R}$,
which leads to crossing between the two bands for $|h|< 2$, band touching for $|h|= 2$, and presence of a band gap for $|h|> 2$. 

The time-dependent equations of the cross-stitch lattice in the presence of onsite nonlinearity read
\begin{equation}
\begin{split}
\hspace{-5mm}
i\dot{a}_n &= -a_{n-1} - a_{n+1} - b_{n-1} - b_{n+1} - h b_{n} +\gamma a_n|a_n|^2 \\
i\dot{b}_n &=  -a_{n-1} -a_{n+1} - b_{n-1} - b_{n+1} - h a_{n} +\gamma b_n|b_n|^2
\end{split}
\label{eq:CS_lin1}
\end{equation}
where $\gamma$ is the nonlinearity strength.
As we have in Sec.\ref{sec:NLFB}, the CLSs of the linear regime can be continued as compact discrete breathers written as Eq.(\ref{eq:FB_states1},\ref{eq:v_FB_states1}) with frequency $\Omega = E_{FB} + \gamma A^2$:
\begin{equation}
\mathcal{C}_{n,n_0} (t)= A
     \begin{pmatrix}
           1 \\
           -1
         \end{pmatrix}
\delta_{n,n_0} e^{-i\Omega t}
\label{eq:CS_CDB1}
\end{equation}
In order to study the linear stability of this model, we linearize Eq.(\ref{eq:CS_lin1}) around the compact discrete breathers Eq.(\ref{eq:CS_CDB1}), and we numerically calculate the eigenenergies of the resulting model for different values of $g = \gamma A^2$ (obtained fixing $A=1$). 

The outcome of our computations can be phrased in the following way. 
Consider first a weakly nonlinear compact discrete breather with $|g| \ll 1$. 
Due to $U=1$ the linear CLS states
are all degenerate but span an orthonormal eigenvector basis of the flat band Hilbert subspace. Therefore, the degeneracy
is harmless, and continuing one CLS into the nonlinear regime will not lead to any resonant interactions with
neighboring CLSs. Therefore, a compact discrete breather whose frequency $\Omega$ is not in resonance with the dispersive part of the linear spectrum $E(q)$ is linearly stable. However, if a compact discrete breather is tuned into
resonance with the dispersive part of the linear spectrum, it will become linearly unstable due to the resonance with
extended dispersive states. 
If we tune the nonlinearity to a finite strength, non-perturbative effects will lead to additional instability windows for
compact discrete breathers.

\begin{figure}[h]
\includegraphics[width=\columnwidth]{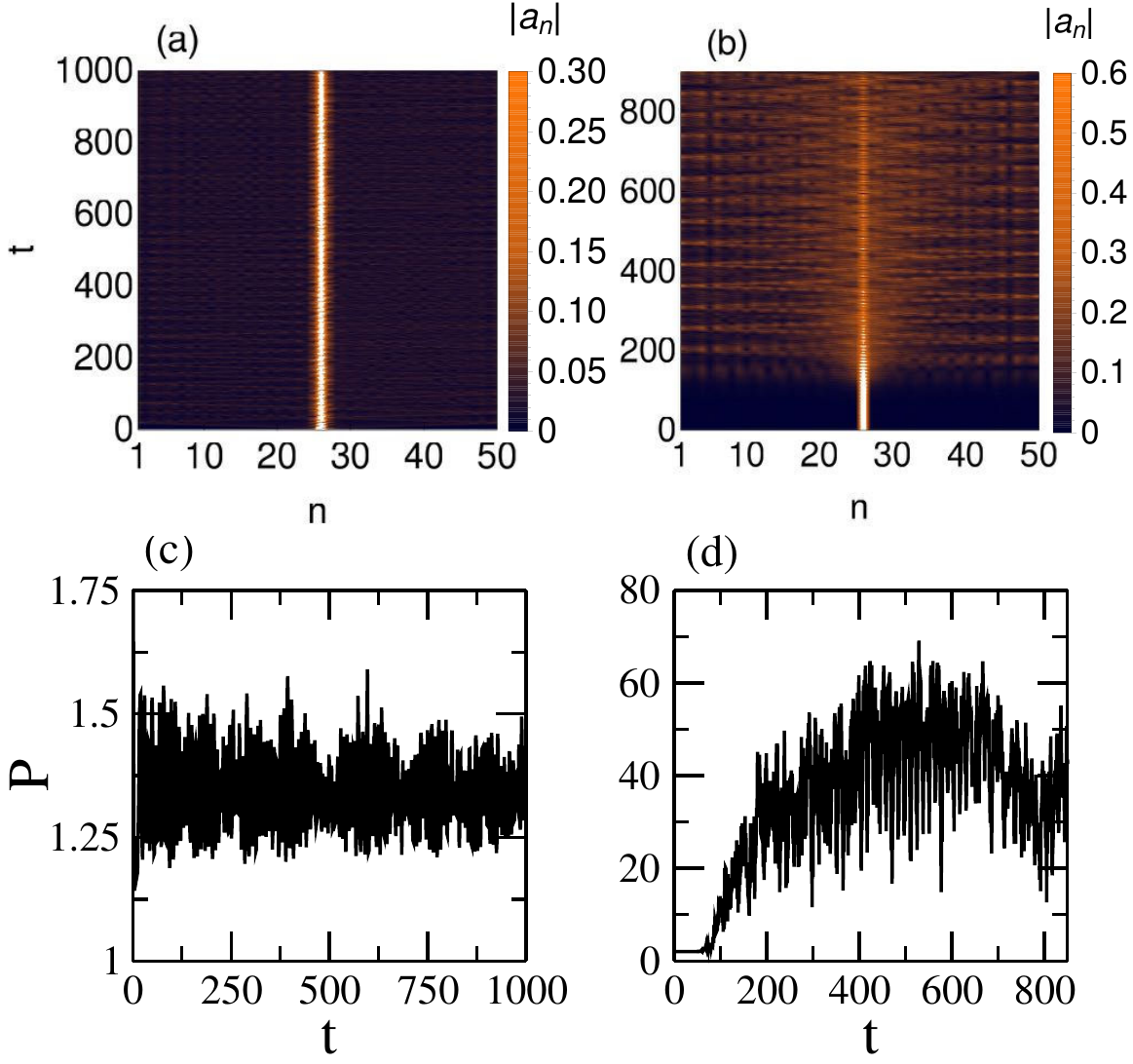}
\caption{Cross-Stitch. (a) and (b): time evolution of the components $a_n(t)$ of an initially perturbed
compact breathers.
(c) and (d): time evolution of the participation number $P$. 
Plots correspond to: $h=3,\,g=5$ (a),(c)
$h=1,\,g=1$ (b),(d). 
}
\label{figcsevol}
\end{figure}

In Fig.\ref{figcsevol}, we show the time evolution of perturbed
compact discrete breathers. We
choose the amplitude of the compact breather to initially be $A=1$, and then we introduce an initial
uniform random perturbation with
maximum amplitude $10^{-3}$ along the whole chain of
$N=50$ unit-cells. In Fig.\ref{figcsevol}(a) and (b) we show the
time evolution of the $|a_n(t)|$ component for $h=3,\,g=5$ and
$h=1,\,g=1$, respectively. Plot (a) has been obtained for $h=3$
and $g=5$, when the frequency $\Omega=3+5=8$ of the compact discrete
breather is located outside the dispersive band $[-7,1]$.
The numerical
simulation of the time-evolution shows stability of the compact
breather. Instead, plot (b) has been obtained for $h=1$ and
$g=1$, when the frequency $\Omega=1+1=2$ of the compact breather is in resonance with the dispersive band
$[-5,3]$. The breather will start to radiate and reduce its amplitude, however the resonance condition will not be destroyed down to
the linear level since the linear flat band is resonating with the dispersive one.
Thus the compact discrete breather is
unstable and will be completely destroyed during its perturbed evolution.
The stable and the unstable behavior of these two cases
are confirmed in Fig.\ref{figcsevol}(c) and (d),
where we show the time evolution of the participation
number $P = 1 / \sum (|a_n|^4 + |b_n|^4)$. The participation number takes values between unity
(obtained for a single site excitation) to the system size (obtained
for uniformly excited states), $P\in[1,N]$, and it estimates the
number of non-negligibly excited sites. Indeed, in plot (c)
which corresponds to the stable compact discrete
breather shown in Fig.\ref{figcsevol}(a), the participation
number $P$ fluctuates around $1.5$, confirming that only few sites are
excited. In plot (d), which corresponds to the
unstable compact discrete breather shown in
Fig.\ref{figcsevol}(b), the participation number $P$ fluctuates
around $40$, confirming the loss of compactness and the
instability of the compact breather.

\subsection*{Saw-tooth}

\begin{figure}[h]
\center\includegraphics [width=\columnwidth]{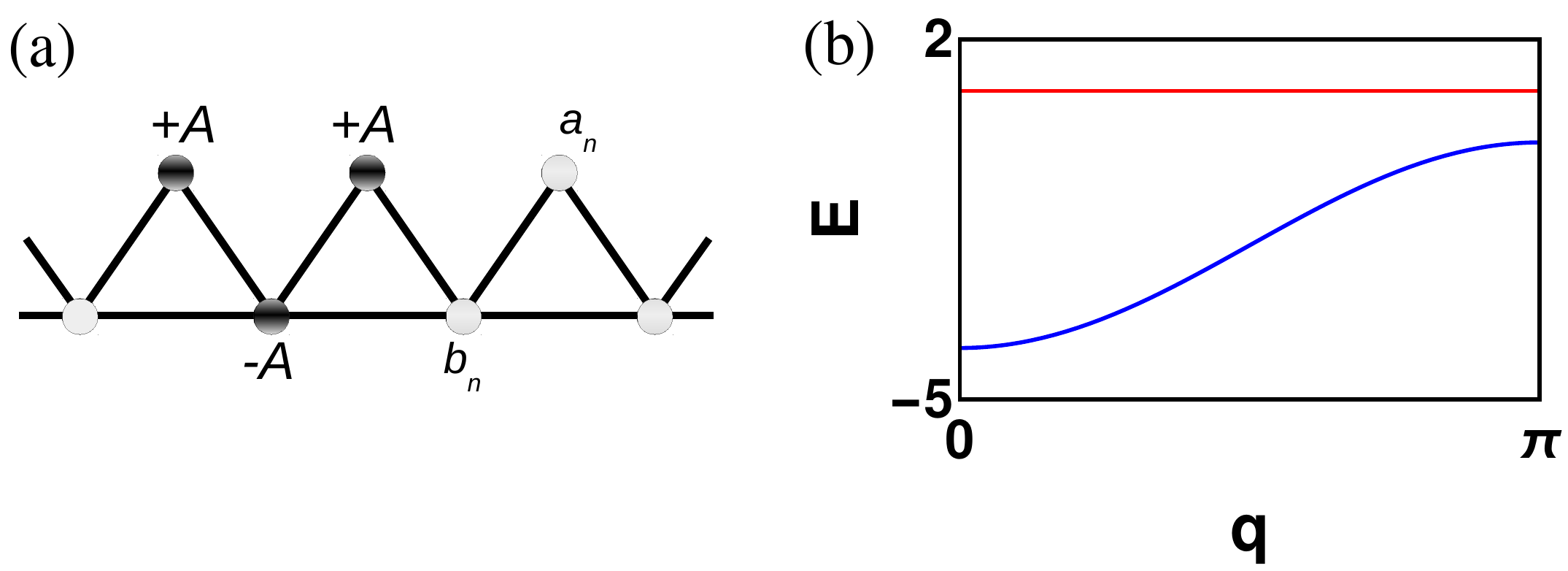}
\caption{(a): profile of the Saw-Tooth lattice.
(b): band structure. } \label{figST1}
\end{figure}

The saw-tooth lattice - Fig.\ref{figST1}(a) - is a one-dimensional
two-band network with one flat band. Associated to the flat band,
there exists a countable set of class $U=2$ compact localized
states, whose homogeneous profile in space is shown by the black
dots in Fig.\ref{figST1}(a). We
recall that in this network, every CLS is non-orthogonal with its
two nearest neighbors. The full band structure of
the model is
\begin{equation}
E_{FB} = 1\ ,\qquad E(q) = -2 - 2\cos(q)
\label{eq:ST_spectrum1}
\end{equation}
which can be observed in Fig.\ref{figST1}(b).
Differently from the cross-stitch case, the spectral bands of this
model cannot be tuned by certain free parameter, and the network
possesses a band gap between the dispersive and the flat
band.

The time-dependent equations Eq.(\ref{eq:FB_ham_NL1}) of the saw-tooth in presence of onsite nonlinearity read
\begin{equation}
\begin{split}
i\dot{a}_n &= - b_{n} - b_{n+1} +\gamma a_n|a_n|^2 \\
i\dot{b}_n &= -b_n - b_{n-1} - b_{n+1} -a_{n-1} -a_{n}+\gamma b_n|b_n|^2
\end{split}
\label{eq:ST_lin1}
\end{equation}
The CLSs of the linear regime can be continued as compact discrete
breathers written Eq.(\ref{eq:FB_states1},\ref{eq:v_FB_states1})
with frequency $\Omega = E_{FB} + \gamma A^2$
\begin{equation}
\mathcal{C}_{n,n_0} (t)= A
\left[
     \begin{pmatrix}
           1 \\
           0
         \end{pmatrix}\delta_{n,n_0-1} +
     \begin{pmatrix}
           1 \\
           -1
         \end{pmatrix}
         \delta_{n,n_0}
         \right]
 e^{-i\Omega t} 
\label{eq:ST_CDB1}
\end{equation}

\begin{figure}[h]
\includegraphics[width=\columnwidth]{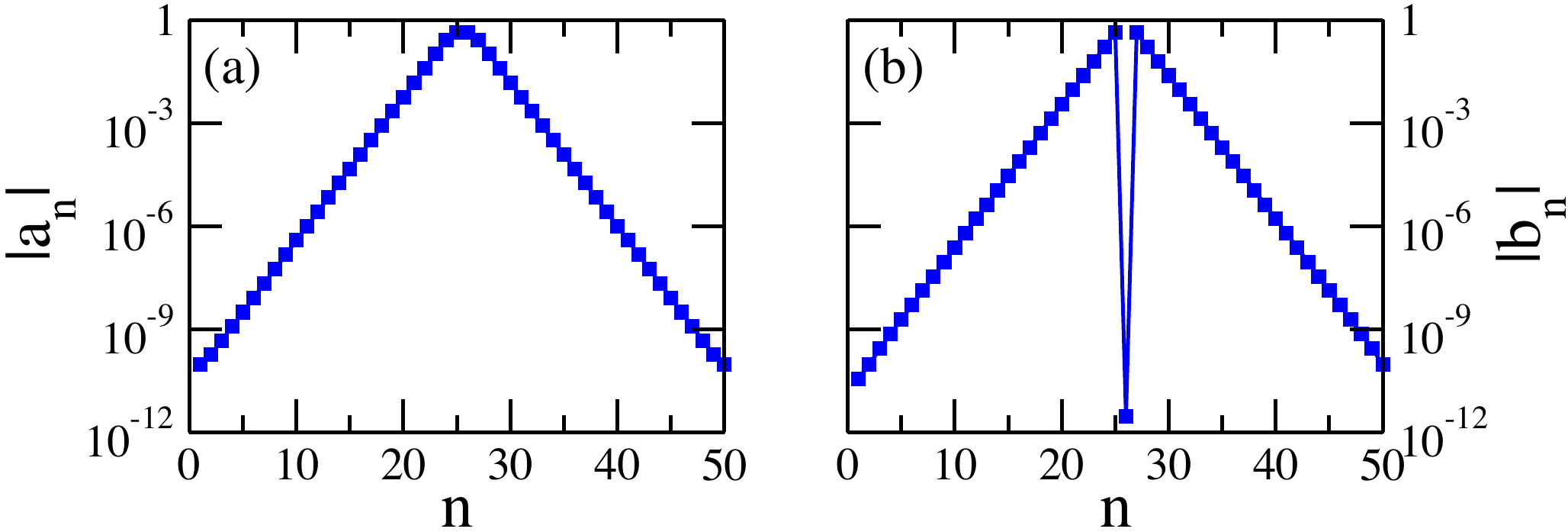}
\caption{Saw-Tooth. $a_n$ component (a) and
$b_n$ component (b) of the unstable eigenvector for $g =-10^{-3}$ and real eigenvalue $EV = 2.987\times
10^{-5}$. } \label{figSTevec}
\end{figure}

Comparing to the $U=1$ case of the cross-stitch lattice, the new feature is the non-orthogonality of neighboring
CLSs at the linear limit. While the flat band is gapped away from the dispersive band, at weak nonlinearities we can
expect a resonant interaction between neighboring CLSs, which may - or may not - lead to model dependent linear local instability.
It turns out that this instability indeed takes place for the saw-tooth chain. 
There exists a narrow region of instability for  $-0.1<g<0$. 
Therefore, the fact that the linear flat band network is
of class $U=2$ makes compact discrete
breathers unstable even in the
presence of a band gap. However, this instability is local, and therefore might not lead to 
to a destruction of the perturbed compact discrete breather, since there is no way to radiate the excitation
to infinity.
In Fig.\ref{figSTevec} we show the $a_n$ components (a) and
the $b_n$ components (b) of the unstable eigenvector with pure real
eigenvalue $EV= 2.987\times 10^{-5}$ obtained for $g=-0.001$.
The eigenvector is exponentially localized.

\begin{figure}[h]
\includegraphics[width=\columnwidth]{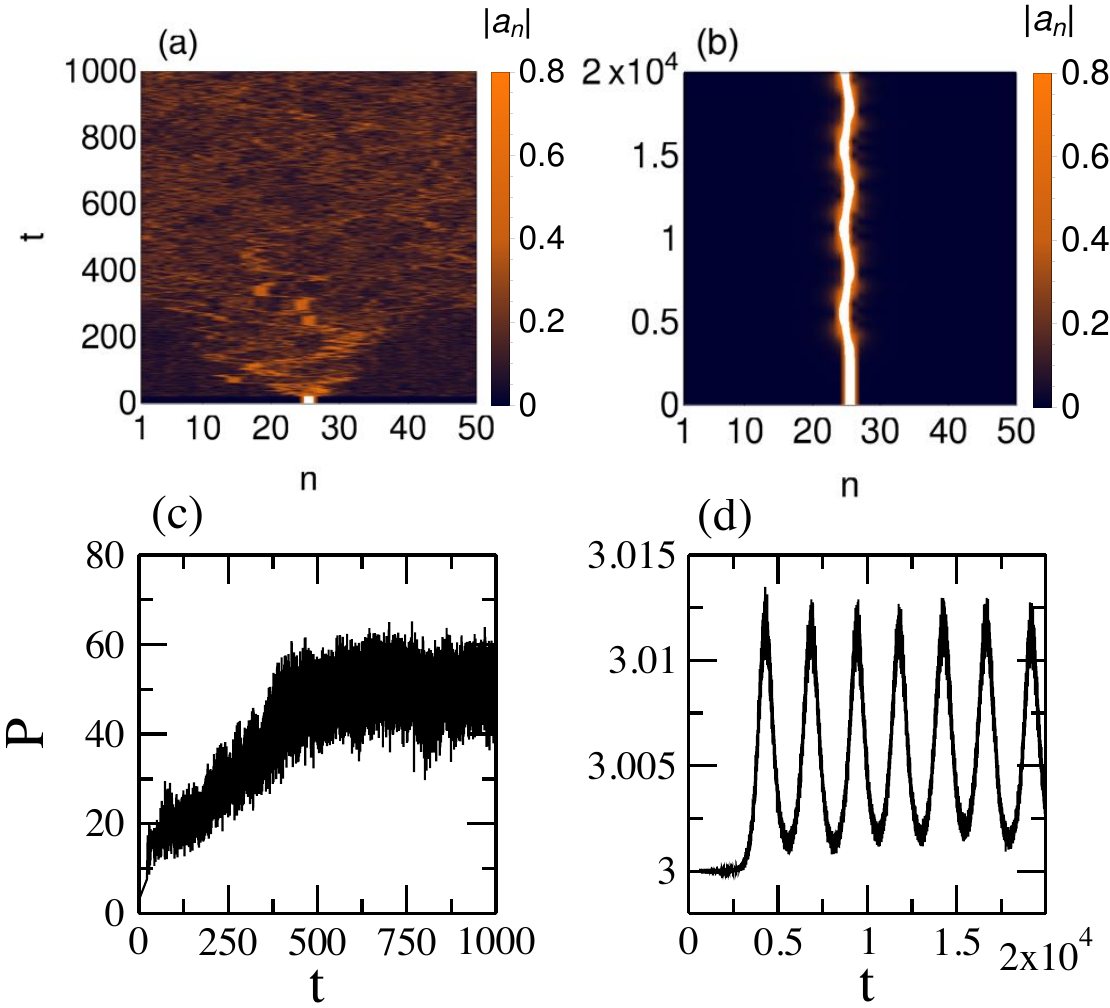}
\caption{Saw-Tooth.
Time evolution of the amplitudes of perturbed
compact breathers (component $|a_n(t)|$) and the participation number. 
Plots correspond to: $g=-1.5$ (a),(c)
$g=-7\times10^{-3}$ (b),(d).
}
\label{figSTevol}
\end{figure}

Let us discuss the time evolution of slightly
perturbed compact discrete breathers, where
a perturbation of order $10^{-3}$ is equidistributed along all the
$N=50$ unit-cells. In Fig.\ref{figSTevol}(a) and (b), we show the time evolution of the $|a_n(t)|$ component for
$g=-1.5$ and $g =-0.007$ respectively, while in
Fig.\ref{figSTevol}(c) and (d) we show the time evolution of the
participation number $P$. In the left column -
Fig.\ref{figSTevol}(a) and (c) correspond to the time
evolution of a compact discrete breather for $g=-1.5$.
In this case, the compact discrete breather is
unstable, since its frequency $\Omega = 1 - 1.5 = -0.5$ is in resonance with the
dispersive band $[-4,0]$. This instability is also
depicted by the participation number $P$. In the right column
 - Fig.\ref{figSTevol}(b) and (d) - we plot the time
evolution of a compact discrete breather for $g=-0.007$
In this case, the 
pure real eigenvalues and the exponentially localized eigenvector yield an oscillatory
behavior in time of the compact discrete breather, which is depicted also by the participation
number $P$.

\section{Conclusions}

In this work, we have discussed the 
properties of compact discrete breathers in some flat band networks.
Linear flat band networks possess compact localized states. In order to continue them into the nonlinear regime
to become compact discrete breathers, a homogeneity condition on the amplitude distribution of CLS has to be
satisfied, which is known to be present for a number of flat band networks. The nonlinear compact discrete breathers
will then persist as compact states, albeit with tuned modified frequencies. 

If these frequencies are in resonance
with dispersive branches of the linear flat band network, then the disrete breather will turn linearly unstable,
which may lead to a complete destruction of the perturbed breather by dissolving it into dispersive states.
If the CLSs form an orthonormal set at the linear limit, no further instabilities are expected in the weakly nonlinear
regime. 
So all it needs to have a stable compact discrete breather at the weakly nonlinear limit, is to tune the flat band energy
out of resonance with the dispersive bands.

However, there exist flat band networks for which the CLSs are not orthogonal. In these cases, the flat band is gapped
away from the dispersive spectrum, and resonances with the dispersive spectrum are avoided in the weakly
nonlinear regime. But the overlap with nearest neighbor CLS states can lead to a local instability in the weakly
nonlinear regime. We indeed observe that this is the case for the saw-tooth chain. Remarkably the instability
does not lead to a complete destruction of the breather, and instead yields a local oscillation of the excitation.

The class of heterogeneous CLSs cannot be continued as
compact discrete breathers. However, as discussed in
\cite{johansson2015compactification}, flat band networks that admit
heterogeneous CLSs in presence of local nonlinearity admit
families of exponentially localized discrete breathers. Additional fine-tuning of parameters and functions 
can lead to a compactification for a countable set of discrete breathers.

\section*{Acknowledgements}
The authors acknowledge financial support from IBS (Project Code No. IBS-R024-D1). 
A.M. acknowledges support from the Ministry of Education and Science of Serbia (Project III45010).

\section*{Appendix}

\appendix

\section{Bloch States Representation}\label{sec:app_1_bloch}

Let us consider Eq.(\ref{eq:FB_ham_NL1}) linearized around one compact discrete breather $C_{n,n_0} (t)$,
for $g \equiv \gamma A^2 $
\begin{equation}
\begin{split}
i \dot{\epsilon}_n &=
  H_0\epsilon_n + H_1\epsilon_{n+1} + H_1^\dagger\epsilon_{n+1} \\
  & + g  \sum_{l=0}^{U-1} \Gamma_l \big(2 \epsilon_{n} + e^{- i 2 \Omega t} \epsilon_{n} ^*   \big) \delta_{n,n_0 + l}
\end{split}
\label{eq:FB_ham_NL3_app}
\end{equation}
where $\{ \Gamma_l \}_{l=0}^{U-1}$ are the projector operators of the vector $\psi_n$ over a compact discrete breather $C_{n,n_0} (t)$
\begin{equation}
 \Gamma_l  = \sum_{j=1}^\nu a_{l,j} {\bf e}_j \otimes {\bf e}_j
\label{eq:projector_app}
\end{equation}
The expansion Eq.(\ref{eq:bloch_expansion}) in Bloch states 
\begin{equation}
\epsilon_n = \frac{1}{\sqrt{N}} \sum_{q} \phi_q e^{iqn}
\label{eq:bloch_expansion_app}
\end{equation}
maps Eq.(\ref{eq:FB_ham_NL3_app}) to
\begin{equation}
\begin{split}
&\qquad i \frac{1}{\sqrt{N}}\frac{\partial}{\partial t} \sum_{q} \phi_q e^{iqn} = 
   \frac{1}{\sqrt{N}}\sum_{q}  H(q)\phi_q e^{iqn} \\
   &+ \frac{ g }{\sqrt{N}} \sum_{q}  \Bigg[ \sum_{l=0}^{U-1} \Gamma_l \big(2 e^{iqn} \phi_q  + e^{- i 2 \Omega t} e^{-iqn} \phi_q^* \big) \delta_{n,n_0 + l} \Bigg] 
\end{split}
\label{eq:FB_ham_bloch1_app}
\end{equation}
where $H(q) \equiv  H_0 + e^{iq}H_1+ e^{-iq}H_1^\dagger$ is the Bloch matrix. This matrix  has $i=1,\dots,\nu$ eigenvalues $\lambda_q^i$ and eigenvectors $v_q^i$, where $\lambda_q^1 = E_{FB}$ and $ v_q^i = w_q$.
Let us multiply Eq.(\ref{eq:FB_ham_bloch1_app}) by $\frac{1}{\sqrt{N}}  e^{ - i\hat{q}n}$ and sum over the lattice $\sum_{n=1}^N$. This yields to
\begin{equation}
\begin{split}
& i \frac{\partial}{\partial t} \sum_{q} \phi_q \left[ \frac{1}{N} \sum_{n=1}^N e^{i(q - \hat{q})n} \right] = 
   \sum_{q}  H(q)\phi_q \left[ \frac{1}{N}\sum_{n=1}^N e^{i(q - \hat{q})n} \right] \\
   &+ \frac{ g }{N}   \sum_{q}  \Bigg[ \sum_{l=0}^{U-1} \Gamma_l \bigg(2 e^{i(q - \hat{q}) (n_0 + l)}  \phi_q \\
   &\qquad \qquad + e^{- i 2 \Omega t} e^{-i (q + \hat{q}) (n_0 + l) }  \phi_q^*  \bigg)  \delta_{n,n_0 + l}  \Bigg]  
\end{split}
\label{eq:FB_ham_bloch1_a_app}
\end{equation}
Without loss of generality, we choose $n_0 = 0$. The relation 
\begin{equation}
 \frac{1}{N} \sum_{q}  e^{iqn} = \delta_{q,0}
\label{eq:bloch_ortho1_app}
\end{equation}
yields to Eq.(\ref{eq:FB_ham_bloch2}) in Sec.\ref{sec:Bloch}
\begin{equation}
\begin{split}
&\qquad \qquad \qquad  i \dot{\phi}_{\hat{q}} =  H(\hat{q})\phi_{\hat{q}} \\
&+ \frac{ g }{N}  \sum_{q}  \Bigg[ \sum_{l=0}^{U-1} e^{-i\hat{q}l} \Gamma_l \bigg(2 e^{iq l} \phi_q + e^{- i 2 \Omega t}  e^{-i q l } \phi_q^*  \bigg)  \Bigg]
\end{split}
\label{eq:FB_ham_bloch2_app}
\end{equation}
Let us now expand $\phi_{\hat{q}} $ in the Bloch eigenbasis 
\begin{equation}
\begin{split}
\phi_{\hat{q}} = f_{\hat{q}} w_{\hat{q}}  +  \sum_{i=2}^\nu d_{\hat{q}}^i v_{\hat{q}}^i 
\end{split}
\label{eq:bloch_exp1_app}
\end{equation}
where $f_q , d_{\hat{q}}^i \in\mathbb{C}$ are time-dependent complex numbers. Eq.(\ref{eq:FB_ham_bloch2_app}) becomes 
\begin{equation}
\begin{split}
& i \frac{\partial}{\partial t} \Big( f_{\hat{q}} w_{\hat{q}} +  \sum_{i=2}^\nu d_{\hat{q}}^i v_{\hat{q}}^i \Big) = 
  E_{FB}  f_{\hat{q}} w_q  +  \sum_{i=2}^\nu \lambda_q^i d_{\hat{q}}^i v_q^i  \\
   &+ \frac{ g }{N}  \sum_{q}  \Bigg\{ \sum_{l=0}^{U-1} e^{-i\hat{q}l} \Gamma_l \bigg[2 e^{iq l} \Big(  f_q w_q  +  \sum_{i=2}^\nu d_q^i v_q^i  \Big) \\
  &\qquad\qquad\quad  + e^{- i 2 \Omega t}  e^{-i q l } \Big(  f_q^* w_q  +  \sum_{i=2}^\nu d_q^{i *} v_q^i  \Big)   \bigg]  \Bigg\}  \ . 
\end{split}
\label{eq:FB_ham_bloch3_app}
\end{equation}
Next, we regroup Eq.(\ref{eq:FB_ham_bloch3_app}) in terms of $f_q $ and $ d_{\hat{q}}^i$ and we multiply it by $w_q^*$. By the orthogonality of the eigenvectors $w_q$ and $v_q^i$ of the Bloch matrix $H(q)$, we obtain Eq.(\ref{eq:FB_ham_bloch_DB}) for the flat band component $f_q $ 
\begin{equation}
\begin{split}
&\qquad\qquad\quad  i \dot{f}_{\hat{q}}   =
  E_{FB}  f_{\hat{q}}  \\
  &+ \frac{ g }{N}  \sum_{q}  \Bigg\{ \sum_{l=0}^{U-1} e^{-i\hat{q}l}
  \Big( 2 e^{iq l} f_q + e^{- i 2 \Omega t}  e^{-i q l } f_q^* \Big) \Gamma_l w_q \cdot w_q^* \\
& + \sum_{l=0}^{U-1}  \Bigg[ \sum_{i=2}^\nu  e^{-i\hat{q}l} \Big( 2 e^{iq l} d_q^i + e^{- i 2 \Omega t}
  e^{-i q l } d_q^{i *} \Big) \Gamma_l v_q^i \cdot w_q^* \Bigg] \Bigg\}
\end{split}
\label{eq:FB_ham_bloch_DB_app}
\end{equation}
Analogously, we obtain Eq.(\ref{eq:FB_ham_bloch_DB2}) for the dispersive bands component $ d_{\hat{q}}^i$ by multiplying  Eq.(\ref{eq:FB_ham_bloch3_app}) by $v_q^{j *}$ and using the orthogonality of the eigenvectors of the Bloch matrix $H(q)$
\begin{equation}
\begin{split}
&\qquad\qquad\quad  \dot{d}_{\hat{q}}^j =
 \lambda_q^j d_{\hat{q}}^j  \\
 &+ \frac{ g }{N}  \sum_{q}  \Bigg\{ \sum_{l=0}^{U-1} e^{-i\hat{q}l} \Big( 2 e^{iq l} f_q + e^{- i 2 \Omega t}  e^{-i q l } f_q^* \Big) \Gamma_l w_q \cdot v_q^{j *} \\
& + \sum_{l=0}^{U-1}  \Bigg[ \sum_{i=2}^\nu  e^{-i\hat{q}l} \Big( 2 e^{iq l} d_q^i + e^{- i 2 \Omega t}  e^{-i q l } d_q^{i *} \Big) \Gamma_l v_q^i \cdot v_q^{j *} \Bigg] \Bigg\}
\end{split}
\label{eq:FB_ham_bloch_DB2_app}
\end{equation}

\section{Strained Coefficient Method}\label{sec:app_2_strained}

Let us consider Eq.(\ref{eq:FB_ham_bloch_DB_U=1}) in Sec.\ref{sec:arnold_tongue_U=1} for a class $U=1$ flat band network
\begin{equation}
\begin{split}
 i \dot{f}_{\hat{q}}  & =
  E_{FB}  f_{\hat{q}}  + \frac{ g }{N}  \sum_{q}  \Big( 2 e^{iq l} f_q + e^{- i 2 \Omega t}  e^{-i q l } f_q^* \Big)  \\
i \dot{d}_{\hat{q}}^j& =
 \lambda_q^j d_{\hat{q}}^j  + \frac{ g }{N} \sum_{i=2}^\nu   \Bigg\{  \sum_{q}
  2 d_q^i + e^{- i 2 \Omega t}  d_q^{i *}  \Bigg\}  \Gamma_0 v_q^i \cdot v_q^{j *}
\end{split}
\label{eq:FB_ham_bloch_DB_U=1_app}
\end{equation}
and the expansion of $d_q$ and $\lambda_{\hat{q}}$ in Eq.(\ref{eq:nayfeh_exp1})
\begin{equation}
\begin{split}
d_{\hat{q}} &=\sum_{k=0}^{+\infty} g^k u_k^{(\hat{q})} \ , \quad
\lambda_{\hat{q}} = m\Omega + \sum_{l=1}^{+\infty} g^l \delta_l  
\end{split}
\label{eq:nayfeh_exp1_app}
\end{equation}
This expansion yields to 
\begin{equation}
\begin{split}
i \frac{\partial}{\partial t} & \sum_{k=0}^{+\infty} g^k u_k^{(\hat{q})}   = 
m\Omega \sum_{k=0}^{+\infty} g^k u_k^{(\hat{q})}  + \sum_{k=0}^{+\infty}  \sum_{l=1}^{+\infty}  g^{k+l}  \delta_l u_k^{(\hat{q})} \\
 &+ \frac{ 1 }{N} \sum_{k=0}^{+\infty}  g^{k+1}  \Bigg\{  \sum_{q} 
 \Big(  2 u_k^{(q)} + e^{- i 2 \Omega t}  u_k^{(q) *} \Big)   \Bigg\}  
\end{split}
\label{eq:FB_ham_bloch_DB_U=1_CS_app}
\end{equation}
Next, we equate the coefficients of each power of $g$ to zero. From $g^0$ we get
\begin{equation}
 i \dot{u}_0^{(\hat{q})}  =  m\Omega  u_0^{(\hat{q})} \quad\Rightarrow\quad u_0^{(\hat{q})}(t) = e^{-im\Omega t} a_0^{(\hat{q})} 
\label{eq:strain_coeff_U=1_g_0_app}
\end{equation}
Without loss of generality, we can assume all the initial conditions to be equal $a_0^{(\hat{q})} \equiv a_0$. For $g^1$ in Eq.(\ref{eq:FB_ham_bloch_DB_U=1_CS_app}) we get
\begin{equation}
 i \dot{u}_1^{(\hat{q})}  =  m\Omega  u_1^{(\hat{q})} + \delta_1 u_0^{(\hat{q})}  
+ \frac{ 1 }{N} \sum_{q}  \Big(  2 u_0^{(q)} + e^{- i 2 \Omega t}  u_0^{(q) *} \Big)  
\label{eq:strain_coeff_U=1_g_1_app}
\end{equation}
which, by Eq.(\ref{eq:strain_coeff_U=1_g_0_app}) reads
\begin{equation}
\begin{split}
& \quad  i \dot{u}_1^{(\hat{q})}  =  m\Omega  u_1^{(\hat{q})} + \\ & \frac{ 1 }{N} \bigg[(N \delta_1 +2) e^{-im\Omega t} a_0 +  e^{- i \Omega(2 - m) t}   a_0 \\
&\qquad \quad +\sum_{q\neq \hat{q} }  \Big(  2 e^{-i \lambda_q t} a_0 + e^{- i (2 \Omega -\lambda_q) t} a_0^*  \Big)  \bigg] 
 \end{split}
\label{eq:bloch_CS_eq_g1_app}
\end{equation}
Using the general solution 
\begin{equation}
y(t) = e^{\int_0^t a(s)ds} + \int_0^t e^{\int_r^t a(s)ds} b(r) dr
\label{eq:sol_diff_eq_app}
\end{equation}
of the first-order differential equation $ \dot{y}(t) = a(t)y(t) + b(t) $, for $m\neq 1$ it follows that 
\begin{equation}
\begin{split}
 u_1^{(\hat{q})}(t) &= e^{-im\Omega t} a_1+ \frac{ 1 }{N} (N \delta_1 +2) a_0 e^{-im\Omega t} t  \\
& + \frac{ i a_0^*}{2 \Omega(1 - m)} \big(e^{- i 2 \Omega t} -  e^{-im\Omega t}\big)    \\
& +  \sum_{q\neq \hat{q} } \bigg[ \frac{i 2  a_0}{ m\Omega - \lambda_q} (e^{-i \lambda_q t} - e^{-im\Omega t}  ) \\
&\qquad \ + \frac{i a_0^*}{ \Omega(2-m) - \lambda_q} \big( e^{-i (2\Omega - \lambda_q )t} - e^{-im\Omega t}   \big)\bigg] 
 \end{split}
\label{eq:bloch_CS_eq_g1_m_neq_1_app}
\end{equation}
To kill the secular term we need to set $\delta_1 = -2 / N$ in the second term of the right hand inside of Eq.(\ref{eq:bloch_CS_eq_g1_m_neq_1_app}). The solution Eq.(\ref{eq:sol_diff_eq_app}) in the case $m=1$ reads
\begin{equation}
\begin{split}
 u_1^{(\hat{q})}(t)&= e^{-i \Omega t} a_1 +  \frac{ 1 }{N}  \big[(N \delta_1 +2) a_0 +   a_0^* \big]  e^{-i \Omega t} t  \\
& +  \sum_{q\neq \hat{q} } \bigg[ \frac{i 2 a_0}{ \Omega - \lambda_q} (e^{-i \lambda_q t} - e^{-i\Omega t}  ) \\
&\qquad \ + \frac{i a_0^*}{ \Omega - \lambda_q} \big( e^{-i (2\Omega - \lambda_q )t} - e^{-i\Omega t}   \big)\bigg] 
 \end{split}
\label{eq:bloch_CS_eq_g1_m_1_app}
\end{equation}
To kill the secular term we need to set 
\begin{equation}
\begin{split}
&\qquad (N \delta_1 +2) a_0+  a_0^* =0  \\
&\quad\Leftrightarrow\quad N \delta_1 = -2 - \frac{a_0^*}{a_0} = -2 + e^{i \theta_0} \\
& \quad\Leftrightarrow\quad \delta_1 =\bigg\{ -\frac{3}{N},-\frac{1}{N} \bigg\}
\end{split}
\label{eq:CS_secular1_app}
\end{equation} 
since $a_0 = I_0 e^{i \theta_0}$, and the coefficients $\{\delta_i\}_i$ in Eq.(\ref{eq:nayfeh_exp1_app}) are real numbers.
The very same procedure discussed above for the components $d_{\hat{q}}$ of the dispersive states can be repeated for the component $f_{\hat{q}}$ of the flat band.
Analogously to the expansion Eq.(\ref{eq:nayfeh_exp1_app}), we perform an expansion for the flat band component $f_{\hat{q}}$ in Eq.(\ref{eq:FB_ham_bloch_DB_U=1_app}) 
\begin{equation}
\begin{split}
f_{\hat{q}} =\sum_{k=0}^{+\infty} g^k v_k^{(\hat{q})}\ ,\qquad E_{FB} = m\Omega + \sum_{l=1}^{+\infty} g^l \sigma_l 
\end{split}
\label{eq:nayfeh_exp_fb1_app}
\end{equation} 
This ultimately lead to the expansion coefficients 
\begin{equation}
\begin{split}
&\sigma_1 = -3,-1  \ ,\quad m=1\\
& \sigma_1 = -2 \ , \qquad \quad m\neq 1
\end{split}
\label{eq:CS_secular1_DB_app}
\end{equation}
The system size $N$ is absent due to the macroscopic degeneracy of the flat band states. Therefore, in Eq.(\ref{eq:bloch_CS_eq_g1_app}) all terms of the sum $\sum_{q\neq \hat{q}}$ have the same time-dependent term $e^{-i E_{FB} t} $.
To kill the secular term in Eq.(\ref{eq:bloch_CS_eq_g1_m_neq_1_app}) for the flat band component $f_q$ for $m\neq 1$, the following condition has to be satisfied 
\begin{equation}
\sigma_1 = - \frac{1}{b_0}  \frac{2}{N }  \sum_{q} b_0 = -2
\label{eq:CS_secular_fano_m_neq_1_app}
\end{equation}
In Eq.(\ref{eq:bloch_CS_eq_g1_m_1_app}) for the flat band component $f_q$ for $m=1$, in order to kill the secular term we need to set 
\begin{equation}
\begin{split}
& \sigma_1  = - \frac{1}{b_0 }  \frac{1}{N}\Bigg[ 2 \sum_{q} b_0+  \sum_{q} b_0^*  \Bigg]
= -2 + e^{i \theta_0} \\
& \quad\Leftrightarrow\quad \sigma_1 =\big\{ -3,-1 \big\}
\end{split}
\label{eq:CS_secular_fano2_app}
\end{equation}
For flat band networks of larger class $U\geq 2$, both expansions Eq.(\ref{eq:nayfeh_exp1_app},\ref{eq:nayfeh_exp_fb1_app}) have to be applied to Eq.(\ref{eq:FB_ham_bloch_DB},\ref{eq:FB_ham_bloch_DB2}). For the saw-tooth case, these equations reduces to 
Eq.(\ref{eq:bloch_U=2_ST_fb},\ref{eq:bloch_U=2_ST_db}) here recalled 
 \begin{equation}
\begin{split}
 i \dot{f}_{\hat{q}}  =
  E_{FB}  f_{\hat{q}}  + \frac{ g }{ \alpha^2 N }  &\sum_{q}  \Bigg\{ (\alpha^2 - 1) \Big(  f_q + e^{- i 2 \Omega t}  f_q^* \Big)   \\
  &+ e^{-i\hat{q}} \Big( 2 e^{iq } f_q + e^{- i 2 \Omega t}  e^{-i q  } f_q^* \Big) \\
  &+    (1 + e^{-iq})  \Big( 2 d_q^i + e^{- i 2 \Omega t}   d_q^{i *} \Big)  \Bigg\}\\
  \end{split}
\label{eq:bloch_U=2_ST_fb_app}
\end{equation}
    \begin{equation}
\begin{split}
i \dot{d}_{\hat{q}} =
 \lambda_q d_{\hat{q}}
 + \frac{ g }{\alpha^2 N} & \sum_{q}  \Bigg\{
    \Big( 2 d_q + e^{- i 2 \Omega t}  d_q^{*} \Big)   \\
&+  e^{-i\hat{q}} \Big( 2 e^{iq } d_q+ e^{- i 2 \Omega t}  e^{-i q  } d_q^{*} \Big)  \\
&+  (1 + e^{iq})  \Big( 2 f_q + e^{- i 2 \Omega t}  f_q^* \Big)
\Bigg\}
\end{split}
\label{eq:bloch_U=2_ST_db_app}
\end{equation}
since in the saw-tooth is a two band problem $\nu = 2$ with the following Bloch vectors $w_q, v_q$ and projector operators $\Gamma_0,\Gamma_1$
\begin{equation}
\begin{split}
& 
w_q=
\frac{1}{\alpha}
 \left(
      \begin{array}{ccc}
         -1  \\
         1 + e^{iq}
      \end{array} \right) 
      \ ,\qquad  \ \ 
      \Gamma_0 = 
       \left(
      \begin{array}{ccc}
         0 & 0 \\
         0 & 1 
      \end{array} \right) 
      \\
      & 
      v_q=
\frac{1}{\alpha}
 \left(
      \begin{array}{ccc}
         1 + e^{-iq}  \\
         1
      \end{array} \right) 
            \ ,\qquad
\Gamma_1 = 
       \left(
      \begin{array}{ccc}
         1 & 0 \\
         0 & 1 
      \end{array} \right) 
\end{split}
\label{eq:ST_bloch}
\end{equation} 
where $\alpha = \sqrt{3 + 2\cos q}$. Eq.(\ref{eq:bloch_U=2_ST_fb_app},\ref{eq:bloch_U=2_ST_db_app}) expanded via Eq.(\ref{eq:nayfeh_exp1_app},\ref{eq:nayfeh_exp_fb1_app}) leads to additional time-periodic terms dependent on the wave number $q$ (called {\it polarized terms}). These terms therefore do not influence the zeroing condition of the secular term presented above for class $U=1$ flat band networks.


\section{Bogoliubov Expansion}\label{sec:app_3_Bogo}

Let us consider Eq.(\ref{eq:FB_ham_bloch_DB_U=1}) for the dispersive component $d_q^i$ for one of the dispersive band $i$ considered for one component only
\begin{equation}
\begin{split}
i \dot{d}_q^i& =
 \lambda_q^i d_q^i  + \frac{ g }{N} 
\left[  2 d_q^i + e^{- i 2 \Omega t}  d_q^{i *} \right]
\end{split}
\label{eq:FB_ham_bloch_DB_U=1_app2}
\end{equation}
Let us simplify the notation, by dropping the $i$. 
We now apply the Bogoliubov expansion to Eq.(\ref{eq:FB_ham_bloch_DB_U=1_app2})
\begin{equation}
d_q  = a_{q}  e^{- i \omega t}  +  b_{q}^*  e^{- i (2\Omega -  \omega) t}\ .    
\label{eq:bogoliubov_tr_app}
\end{equation}
This yields to 
\begin{equation}
\begin{split}
&\qquad \omega a_{q} e^{- i \omega t}  + (2\Omega -  \omega)  b_{q}^*  e^{- i (2\Omega -  \omega) t}  \\
&\qquad=   \lambda_{q}  a_{q} e^{- i \omega t}  +  \lambda_{q}  b_{q}^*  e^{- i (2\Omega -  \omega) t} \\
&+ \frac{ g }{N}   \Big[ \big(2 a_{q}  +  b_{q} \big) e^{- i \omega t} + \big(2 b_{q}^*  +  a_{q}^*  \big) e^{- i (2\Omega -  \omega) t}   \Big]  
\end{split}
\label{eq:Bog_exp_1_app}
\end{equation}
which can be decoupled in two equations 
\begin{equation}
\begin{split}
&\omega a_{q} =  \lambda_{q}  a_{q} + \frac{ g }{N}  \big(2 a_{q}  +  b_{q} \big) \\
 (2\Omega & -  \omega)   b_{q}^* = \lambda_{q}  b_{q}^* +  \frac{ g }{N}  \big(2 b_{q}^*  +  a_{q}^*  \big)
\end{split}
\label{eq:Bog_exp_2_app}
\end{equation}
In matrix form, Eq.(\ref{eq:Bog_exp_2_app}) reads
\begin{equation}
\begin{split}
\omega \left(
      \begin{array}{ccc}
        a_{q}   \\
        b_{q}  
      \end{array} \right)
&=
\left(
      \begin{array}{ccc}
         \tilde{\lambda} &   \epsilon   \\
      - \epsilon  &   2\Omega    -  \tilde{\lambda} 
      \end{array} \right)
\left(
      \begin{array}{ccc}
        a_{q}   \\
        b_{q}  
      \end{array} \right)
\end{split}
\label{eq:Bog_exp_3_app}
\end{equation}
for $ \tilde{\lambda} = \lambda_{q} + 2 g / N$ and $\epsilon =  g / N$. The eigenvalues of this system
\begin{equation}
\begin{split} 
\omega_{1,2}  
&= \Omega \pm \sqrt{  \Omega^2 - 2 \tilde{\lambda} \Omega + \tilde{\lambda}^2  - \epsilon^2  } \\
\end{split}
\label{eq:Bog_exp_eigenval_app}
\end{equation}
The argument of the square root is equal to zero when 
\begin{equation}
\begin{split}
\Omega_{1,2} &=  \frac{ 2 \tilde{\lambda} \pm \sqrt{  4 \tilde{\lambda}^2 - 4(  \tilde{\lambda}^2  - \epsilon^2 ) } }{2} 
= \tilde{\lambda} \pm    \epsilon 
\end{split}
\label{eq:Bog_cond1_app}
\end{equation}
This yields complex eigenvalues $\omega$ in Eq.(\ref{eq:Bog_exp_1_app}) for $ \tilde{\lambda} -  \epsilon \leq \Omega \leq  \tilde{\lambda} +  \epsilon$ which translates into 
\begin{equation}
\begin{split}
\lambda_q + \frac{ g }{N}    \leq \Omega \leq \lambda_q + \frac{ 3 g }{N}  
\end{split}
\label{eq:Bog_cond2_app}
\end{equation}
which is the Arnold tongue obtained in Eq.(\ref{eq:CS_secular1_DB}).

\section{Numerical Methods}\label{sec:app_4_nummeth}

In the linear stability analysis of the compact discrete breathers, the nonlinear model Eq.(\ref{eq:FB_ham_NL1}) is reduced to
the eigenvalue problem Eq.(\ref{eq:FB_ham_NL3}) of small perturbation $\epsilon_n$ added to a compact discrete breathers $C_{n,n_0}(t)$.
We solve this problem numerically by applying an IMSL Fortran 
routine called DEVCRG (see \cite{numerics2003imsl} for details). 
The time evolution of the perturbed compact discrete breathers has been obtained by direct integration (for example
\cite{gligoric2016nonlinear}). These numerical simulations have been performed using a $6$th order Runge-Kutta
procedure.

\bibliographystyle{apsrev4}
\let\itshape\upshape

\bibliography{flatband}

\end{document}